\documentclass[%
reprint,
superscriptaddress,
amsmath,amssymb,
aps,
prmaterials,
]{revtex4-2}

\usepackage{graphicx}
\usepackage{dcolumn}
\usepackage{bm}
\usepackage{hyperref}

\usepackage{cleveref}
\usepackage{siunitx}
\usepackage{mathtools}
\usepackage[caption=false]{subfig}
\usepackage{enumitem}
\usepackage{multirow}
\usepackage{scalerel,amssymb}
\usepackage{verbatim}
\usepackage{xcolor}
\newcommand{\etal}{\textit{et al.} }

\UseRawInputEncoding

\begin{document}
\title{Critical Peeling of Tethered Nanoribbons}

\author{Andrea Silva}
\email{ansilva@sissa.it}
\affiliation{CNR-IOM, Consiglio Nazionale delle Ricerche - Istituto Officina dei Materiali, c/o SISSA Via Bonomea 265, 34136 Trieste, Italy}
\affiliation{International School for Advanced Studies (SISSA), Via Bonomea 265, 34136 Trieste, Italy.}

\author{Erio Tosatti}
\affiliation{International School for Advanced Studies (SISSA), Via Bonomea 265, 34136 Trieste, Italy.}
\affiliation{The Abdus Salam International Centre for Theoretical Physics (ICTP), Strada Costiera 11, 34151 Trieste, Italy}
\affiliation{CNR-IOM, Consiglio Nazionale delle Ricerche - Istituto Officina dei Materiali, c/o SISSA Via Bonomea 265, 34136 Trieste, Italy}

\author{Andrea Vanossi}
\affiliation{CNR-IOM, Consiglio Nazionale delle Ricerche - Istituto Officina dei Materiali, c/o SISSA Via Bonomea 265, 34136 Trieste, Italy}
\affiliation{International School for Advanced Studies (SISSA), Via Bonomea 265, 34136 Trieste, Italy.}


\begin{abstract}
The peeling of an immobile adsorbed membrane is a well known problem in engineering and macroscopic tribology.
In the classic setup, picking up at one extreme and pulling off results in a peeling force that is a decreasing function of the pickup angle.
As one end is lifted, the detachment front retracts to meet the immobile tail.
At the nanoscale, interesting situations arise with the peeling of graphene nanoribbons (GNRs) on gold, as realized, e.g., by atomic force microscopy.
The nanosized system shows a constant-force steady peeling regime, where  the tip lifting $h$ produces no retraction of the ribbon detachment point,  and just an advancement $h$ of the free tail end.
This is opposite to the classic case, where the detachment point retracts and the tail end stands still.
Here we characterise, by analytical modeling and numerical simulations, a third, experimentally relevant, setup where the nanoribbon, albeit structurally lubric, does not have a freely moving tail end, which is instead elastically tethered.
Surprisingly, novel nontrivial scaling exponents appear that regulate the peeling evolution.
As the detachment front retracts and the tethered tail is stretched, power laws of $h$ characterize the shrinking of the adhered length the growth of peeling force and the peeling angle.
These exponents precede the final total detachment as a critical point, where the entire ribbon eventually hangs suspended between the tip and tethering spring.
These analytical predictions are confirmed by realistic MD simulations, retaining the full atomistic description, also confirming their survival at finite experimental temperatures.

\end{abstract}

\maketitle
%
\section{Introduction}
Peeling of adsorbed films stripped off adhesive substrates is a classic subject in tribology and mechanics.
Kendall \cite{Kendall1971,Kendall1975} described theoretically and experimentally the peeling evolution of an immobile adsorbed film or ribbon, once picked up at one extreme and pulled off by a constant force.
The peeling force results, reasonably, a decreasing function of the pickup angle from the parallel to the perpendicular direction.
For arbitrary pickup angles the film detachment point, determined by the  force, moves backward, while the remaining film body and tail end are by construction immobile.

With the advent of nanophysics,  more interesting  situations arise  with the peeling of adsorbed nanostructures.
Peeling of soft DNA strands suggested insights on how the dragged macromolecule behaves and helped characterise its mechanical response\cite{Manohar2008,Vilhena2018}.
Another case of interest is the peeling, via atomic force microscopy (AFM), of
graphene nanoribbons (GNRs) initially adsorbed on gold (111) surfaces \cite{Kawai2016,Pawlak2020,Gigli2019a}.
In these experiments one GNR end is picked up by a nanotip and lifted vertically, gradually stripping off the full nanoribbon.
The main new feature in this case is represented by the lower corrugation of the interface, together with a high in-plane ribbon stiffness, which permits some level of sliding  on the substrate.
In fact, an infinite physisorbed 2D graphene sheet forms with the perfect close-packed metal surface an incommensurate interface,  that would typically exhibit structural lubricity (“superlubricity”) \cite{Vanossi2020a} -- a state where static friction is zero, so that any pulling force would cause sliding in the first place.
An adsorbed nanoribbon of finite length, even if still incommensurate and forming  the typical 1D moir\'e with the underlying gold surface \cite{Gigli2017}, is atomically pinned at least at head and tail.
In principle, despite the intrinsic incommensurability, even its full length could not be structurally lubric, but pinned to the underneath substrate \cite{Vanossi2013}.
Modest as they may be, these  pinning  sources turn the pure
stripping by a lifting tip of a nanoribbon, even a superlubric one, into a combination of  peeling and sliding, the latter characterized by atomistic stick-slip\cite{Gigli2017,Gigli2018,Gigli2019}.

A counterpart to Kendall's peeling theory designed for superlubric GNRs, mathematically transparent under the simplifying assumptions of inextensibility and of zero corrugations (therefore without stick-slip) was recently put forward by Gigli \etal  for a typical AFM setup \cite{Gigli2019}.
In perpendicular peeling, the configuration of the ribbon is described the by the peeling angle $\theta$, defined as the angle between the adsorbed and lifted segments, and bending curvature $1/R$.
The evolution of ribbon shape during peeling is obtained numerically as a function of lifting height $h$, highlighting an interesting initial stage, where peeling angle and detachment curvature display a nontrivial growth, followed by a steady peeling regime, where $\theta$ has reached $\pi/2$ and the curvature is fixed.
In that regime, a lifting amount $h$ produces no retraction of the ribbon detachment point but a simple advancement of the free tail end (and therefore a decrease of the ribbon-surface adhered contact length) by exactly the lifting value $h$, if one ignores a small stretching of the very stiff graphene lattice.
That is the exact opposite of Kendall’s limit, where the film sticks to the substrate and the detachment point retracts while the tail end stands still.

A third intriguing situation, experimentally relevant, arises when the nanoribbon being picked up and lifted for peeling, albeit structurally lubric and altogether similar to those just described, does not have a freely moving tail.
As it happens, the tail end is in this case trapped-tethered - by some other adsorbate or defect acting as an elastic constraint.
Under this impeding constraint the peeling evolution should change with respect to the two limits described above.
In the example of perpendicular peeling, where one end is vertically lifted by $h$, the tethered tail end is no longer free to slide forward by $h$, despite superlubricity of the adhered nanoribbon.
An elastically tethered tail will yield to some extent, but clearly in the steady state lifting regime the detachment front is forced to retract toward the tail.
What could be reasonably expected here? A first guess could be some relatively uninteresting, parameter-dependent, intermediate regime between the two described above.

The surprising result which we report here, analytical and verified by realistic simulations, is that new nontrivial exponents appear, unrelated to the more regular limiting cases.
The quantitative result depends on the precise stress-strain characteristics of the tethering spring.
The example which we shall demonstrate here is the simplest one, namely a perfectly harmonic spring.
As the tip is lifted, the zero-temperature peeling force of the tethered nanoribbon is predicted to increase as $h^{1/3}$, and the lifting angle to drop
asymptotically as $h^{-1/3}$.
While the detachment front retracts and the tethered tail advances, the adsorbed fraction therefore shrinks as $h^{4/3}$.
A finite GNR length $L$ will of course truncate the power law evolution, which nonetheless prepares for the ideal $L\to\infty$ critical limit.

We will first formulate the analytical energy of a harmonically tethered idealized GNR as a function of the peeling height $h$, whose minimization predicts the evolution of all other variables.
The critical exponents are obtained analytically in the $h\to\infty$ limit.
That result is then confirmed by comparison with  a realistic molecular dynamics simulation of the peeling of a tail-tethered GNR off a (111) gold surface.
While a comprehensive finite-temperature treatment is deferred to a subsequent study, we show that extension of the molecular dynamics simulation to room temperature does not qualitatively destroy our zero temperature predictions, whose experimental verification should be entirely possible.

\section{Results}
\begin{figure}
    \centering
    \includegraphics[width=\columnwidth]{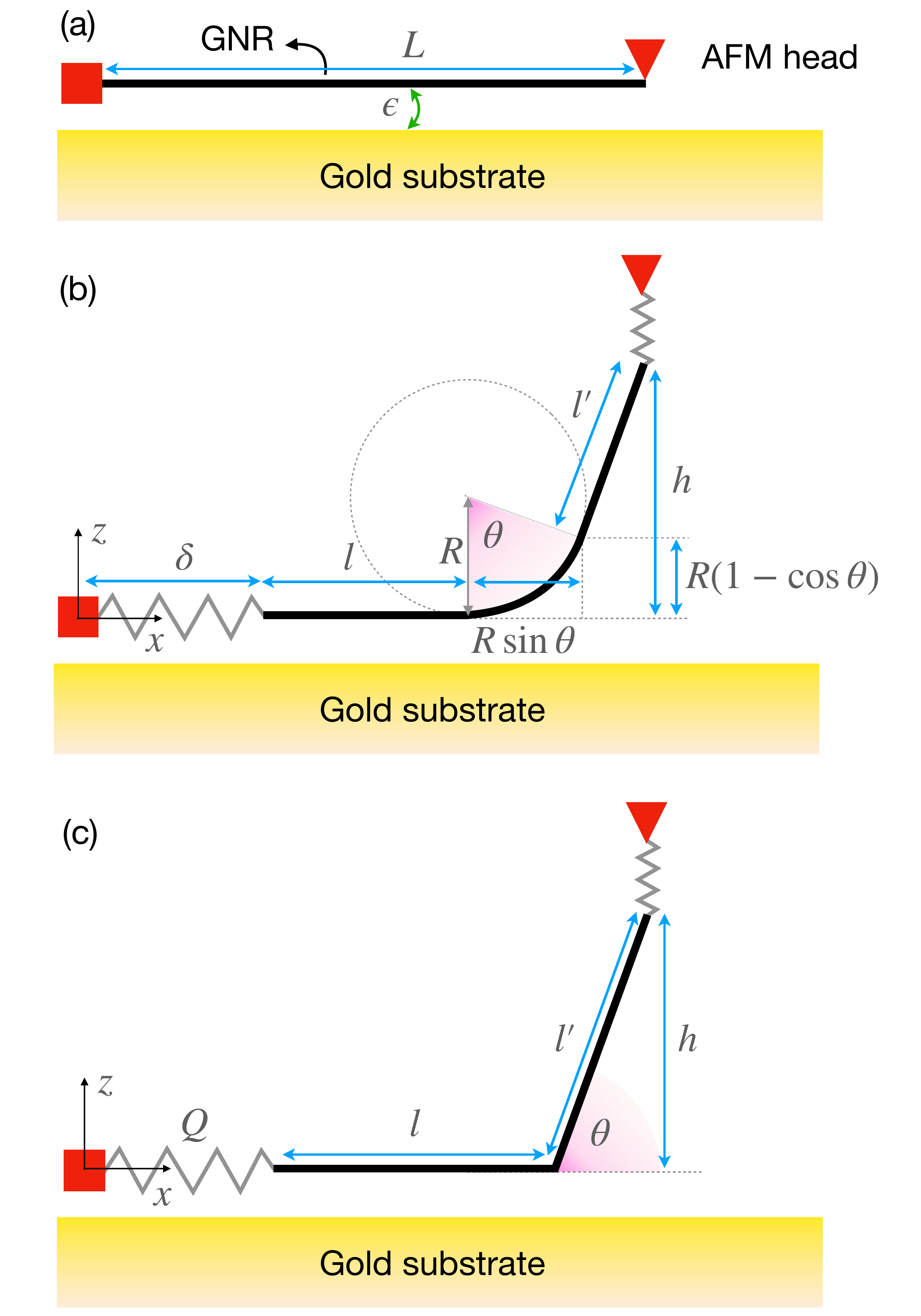}
    \caption{
    Sketch of AFM peeling of a tethered GNR.
    The tip is perpendicularly lifted at height $h$ along $z$ via an infinitely strong tip spring.
    The tail is tethered to the initial position at $x=0$ via a spring $Q$ and physisorbed to the substrate with energy $\epsilon$ per unit length.
    (a) Initial configuration with fully adsorbed GNR along $x$ (the real GNR extends in the $y$ direction for about 7 $\AA$).
    (b) Mid-peeling configuration with bending.
    (c) Mid-peeling configuration without bending.
    }
    \label{fig:sketch-bend}
\end{figure}
To model the tether GNR system, we consider the setup in \cref{fig:sketch-bend}.
This system is in between the known models of Kendall and Gigli, as derived in Section I of the SI.
Note that the GNR is a two dimensional membrane extending in $x$ and $y$.
Here we reduce the system to a 1D model by integrating over $y$ coordinate.
The lifting point is fixed in ($x,y$) mimicking an AFM setup, the adsorbate is free to slide without barrier, as in the superlubric GNR/Au(111) experimental interface, but the tail is tethered.
At the starting configuration $h=0$, the ribbon is flat on the substrate and the tail spring is at rest, $\delta=0$, as sketched in \cref{fig:sketch-bend}a.
The tail end is anchored to a fixed point by a harmonic spring of stiffness $Q$.
As the tip is lifted vertically and peeling proceeds as sketched in \cref{fig:sketch-bend}b, the potential energy comprises four terms: the adhesion energy proportional to the adhered fraction $l$, the elastic energy due to the elongation of the tail spring of stiffness $Q$, the bending energy due to the peeling front, arching at an angle $\theta$ and curvature $R$, and the cost of stretching a GNR of stiffness $K$.
The elongation $\delta$ of the tail spring is expressed as the difference between the total rest length $L$ and the projection onto the horizontal axis of the GNR (see \cref{fig:sketch-bend}b):
\begin{align}
    \delta = L - l - \frac{h}{\tan\theta} - R \tan\frac{\theta}{2}.
\end{align}
Referring to the geometry in \cref{fig:sketch-bend}b, the energy  of the system $E(l, \theta, R; h)$ is
\begin{align}
\label{eq:full_energy}
    E = -\epsilon w l + \frac{Q}{2} \left(L - l - \frac{h}{\tan\theta} - R \tan\frac{\theta}{2} \right)^2 + \nonumber \\
    \frac{K}{2} \left( L - l - \frac{h}{\sin\theta} - R \left(\theta - \tan\frac{\theta}{2}\right) \right)^2 + B w \frac{\theta}{2R},
\end{align}
where $w$ is the width of the GNR, $Q$ is the tail spring constant, $\epsilon$ the adhesion energy per unit area, $B$ the ribbon's bending rigidity and $K$ its effective elastic constant.
The energy in \cref{eq:full_energy} depends parametrically on the tip height $h$.

The parameter values are chosen to mimic real GNR of width $w=\SI{7}{\AA}$ on Au(111), as in AFM experiments\cite{Gigli2019a}: $B \approx \SI{1.2}{eV}$ and $\epsilon\approx\SI{0.017}{eV/\AA^2}$.
Here we focus on a length $L=\SI{30}{nm}$, in a similar range to experimental and simulated values, while numerical solutions for longer GNR can be found in Section II.B of the SI.
The 2D Young modulus of graphene $Y_\mathrm{2D}=\SI{366.2}{eV/\AA^2}$\cite{Bosak2007G_elasticExp,Andrew2012G_elastic} yields an effective stiffness $K=Y_\mathrm{2D} w/L=\SI{0.533}{eV/\AA^2}$.

The tail tethering parameters will depend from the precise occasional nature of the actual constraint found in experiments.
To the best of our knowledge, no systematic characterization of a tethered GNR-like system is present in the literature.
Tentatively, we assume a harmonic constraint and explore the behaviour of the system as a function of the spring constant over several orders of magnitude.
Finally, any  locally stable tethers must  behave as harmonic at least in the beginning of the peeling.

In order to understand the physics of tethered peeling, we focus separately on the beginning and steady state of the peeling, which can be solved analytically.

\subsection{Onset of peeling: in-extensible, bendable GNR}
At the beginning, in the initial peeling transient, the GNR is close to its relaxed length, thus it can be considered as inextensible, $K = \infty$.
The intrinsic elasticity term in \cref{eq:full_energy} reduces in this limit to a Lagrange multiplier, and the adsorbed fraction can be expressed as a function of the peeling angle $\theta$ and bending radius $R$:
\begin{equation}
    l = L - \frac{h}{\sin\theta} - R \left(\theta - \tan\frac{\theta}{2}\right) \label{eq:Lagran}
\end{equation}
As the AFM tip is macroscopically larger than the GNR and the lifting proceeds slower than the relaxation time of the GNR, we consider the lifting point at a fix height $h$ while the GNR relaxes to mechanical equilibrium\cite{Gigli2019}.
The energy is expressed as:
\begin{align}
\label{eq:en_bend}
    E(\theta, R; h) = -L \epsilon  + \epsilon \left( \frac{h}{\sin\theta} + R (\theta-\tan\frac{\theta}{2})\right) \nonumber \\
    + \frac{Q}{2} \left( h \tan\frac{\theta}{2} + R (\theta - 2 \tan\frac{\theta}{2}  \right)^2 + B \frac{\theta}{2R}
\end{align}

At mechanical equilibrium, variations of the energy with respect to all degrees of freedom must vanish
\begin{widetext}
\begin{align}
\label{eq:dFdR}
    \frac{\partial E}{\partial R} &= 0 = R^2 \left[ w \epsilon \frac{\theta-\tan\frac{\theta}{2} }{\alpha(\theta)} + Q h \frac{\tan\frac{\theta}{2}}{\alpha(\theta)} + Q R \right] - \frac{w B\theta}{2\alpha(\theta)}\\
\label{eq:dFdtheta}
    \frac{\partial E}{\partial \theta} &= 0 = w\epsilon\cos\theta\left( \frac{R}{\cos\theta+1} - \frac{h}{\sin^2\theta}\right) + \frac{Q}{\cos\theta+1}(h\tan\frac{\theta}{2} + R \alpha(\theta))(h+R(\cos\theta-1)) + \frac{B w}{2R} ,
\end{align}
\end{widetext}
with the shorthand $\alpha(\theta)=\theta-2\tan\frac{\theta}{2} $.
Solution of \cref{eq:dFdR,eq:dFdtheta} can be computed numerically.

\begin{figure}
    \centering
    \includegraphics[width=\columnwidth]{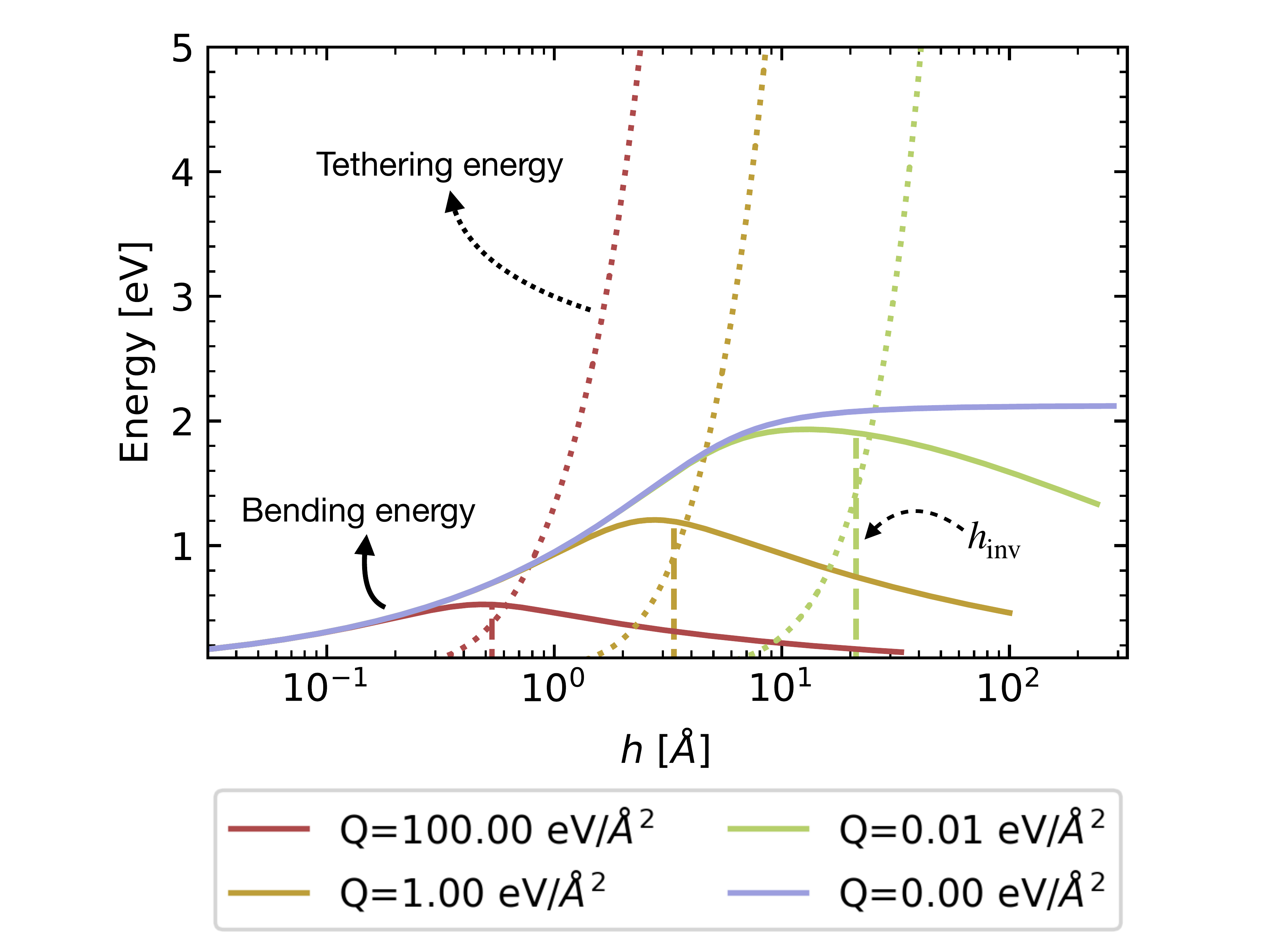}
    \caption{\label{fig:bending_en}
    Bending energy (solid lines) and tethering energy (dotted lines) in the  inextensible model ($K \to \infty$) described by \cref{eq:en_bend}.
    Different colors refer to different values of the tethering spring stiffness $Q$, as reported in the caption.
    The vertical dashed lines of matching colors indicate the inversion height $h_\mathrm{inv}$, see \cref{eq:h_inv}, at which the system switches from bending to tethering regime.
    The energy scale is clipped to focus on the competition between the two terms, the bending one dominating for small $h$, and the tethering one for large $h$.
    }
\end{figure}
It is instructive to decompose the energy into the three contributions appearing in \cref{eq:en_bend}: bending $E_\mathrm{bend}=Bw\theta/2R$, adsorption $E_\mathrm{ads}=-\epsilon w l$ and tethering energy $E_\mathrm{teth}=Q \delta^2/2$.
The relevant contribution we focus on here is the bending and tethering energies while the adsorption contribution and the total energy are reported in Section II.A of the SI.
\Cref{fig:bending_en} shows the evolution of the bending energy $E_\mathrm{bend}$ (solid lines) and tethering energy $E_\mathrm{tether}$ (dotted lines) as a function of the AFM tip height $h$.
The colors refer to different values of the tethering stiffness $Q$, as reported in the legend below the figure.
The bending energy (solid curves) for all tethering strengths except for $Q=0$ first rise at small $h$, reach a maximum, approximately marked by the color-matching dashed line, and then decay.
As the peeling begins, for small $h$, the tethering spring remains almost at rest: $E_\mathrm{thet}$ is almost flat while $E_\mathrm{bend}$ rises, due to the price of building a peeling angle at the peeling front.
The rise continues until the peeling angle gets large enough to demand some advancement of the tethered tail.
At this point (dashed lines \cref{fig:bending_en}) the tail spring starts to load and hinder further sliding of the GNR: $E_\mathrm{tether}$ raises while $E_\mathrm{bend}$ starts to decrease.
As the GNR is disputed between the AFM tip and the tethered tail, the decrease in $E_\mathrm{bend}=\frac{B w \theta}{2R}$ may arise from different mechanisms: either the peeling angle $\theta$ could decrease at constant curvature $R$, or the curvature $R$ might increase at constant angle, or in effect a combination of these two.
After this initial bending-tethering interplay, the bending energy becomes negligible and the peeling enters a steady-state regime dominated by the interplay between the tail and the peeling front, governed by the adhesive energy and the tethering spring extension, as described in the following section.

\begin{figure*}[ht]
    \centering
    \includegraphics[width=0.9\textwidth]{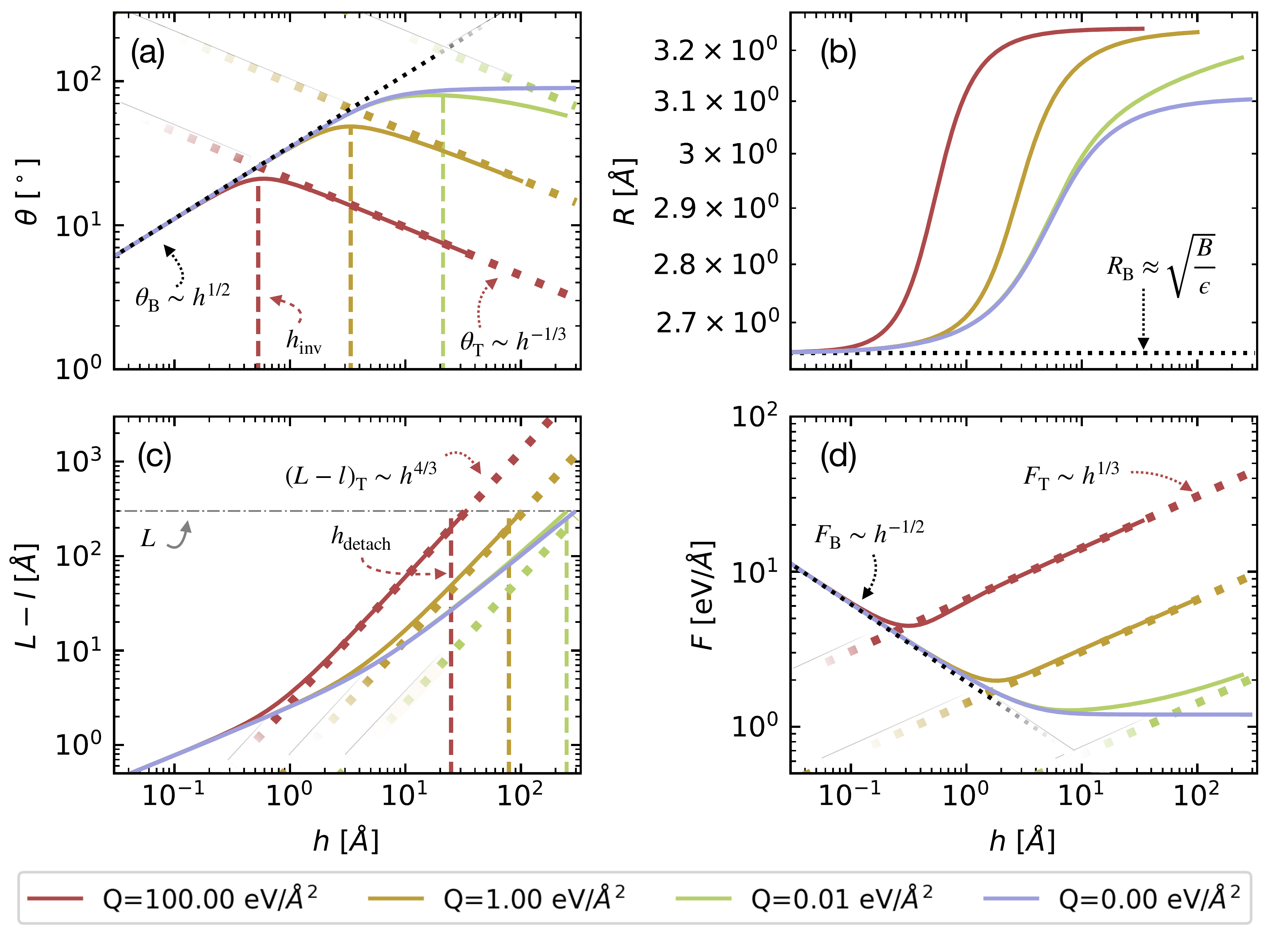}
    \caption{
    Peeling angle $\theta$ (a), bending curvature $R$ (b), detached fraction $L-l$ (c) and force $F$ (d) as a function of tip height $h$.
    Colors refer to different tail spring $Q$, as reported in the legend at the bottom.
    The solid lines in each plot refer to numerical solution of \cref{eq:dFdR,eq:dFdtheta}.
    Black dotted lines in a,b,d report the scaling behaviour in the limit $h\to 0$, shown in \cref{eq:Blimth,eq:BlimR,eq:TlimF}.
    Colored dotted lines in a,c,d are the scaling behaviours in the limit $h\to\infty$, shown in \cref{eq:Tlimth,eq:Tliml,eq:TlimF}.
    }
    \label{fig:model_scaling}
\end{figure*}
The mechanics underpinning the peeling is revealed by the evolution of the GNR configuration.
\Cref{fig:model_scaling}a-c reports the behaviour of the peeling angle $\theta$, the bending curvature $R$, and the detached fraction $L-l$ as a function of the tip height $h$.
\Cref{fig:model_scaling}d shows the value of the force acting on the tip $F=\partial E / \partial h$, which is related to the cantilever frequency shift measurable in AFM experiments\cite{Gigli2019a}.
The initial bending and subsequent tethering regimes are clearly seen in the evolution of the peeling angle in \cref{fig:model_scaling}a, which increases at the beginning of peeling, peaks during the competition between bending and tethering and then starts to decrease.
The bending curvature in \cref{fig:model_scaling}b saturates after the initial stage.
Hence, the decrease in the bending energy $E_\mathrm{bend}$ in \cref{fig:bending_en} is mostly due to the evolution of the detachment angle $\theta$.
For vanishing small tethering $Q\to 0$, the peeling angle approaches $\pi/2$ and the Gigli model is recovered: the peeling proceeds at constant angle and force (blue lines in \cref{fig:model_scaling}a,d), with the GNR unwinding right below the tip.
Coherently with the angle evolution, the crossover heigth separating bending and tethering regimes decreases as the tethering stiffness $Q$ increases.

By considering the limit of small height $h \to 0$ in \cref{eq:dFdR,eq:dFdtheta}, we obtain an analytical approximation of the GNR configuration in the bending limit
\begin{align}
    R_\mathrm{B} &\approx \sqrt{\frac{B}{\epsilon}} \label{eq:BlimR}\\
    \theta_\mathrm{B} &\approx \sqrt{h R_\mathrm{B}} \label{eq:Blimth}\\
    F_\mathrm{B} &\approx \epsilon \sqrt{\frac{R_\mathrm{B}}{h}} . \label{eq:BlimF}
\end{align}
These limits, where the subscript B denotes the bending regime, are shown in \cref{fig:model_scaling}a,b,d in dotted black lines.

\subsection{Steady state peeling: harmonically tethered GNR}
As the peeling enters the steady-state regime and the tail spring begins to load, the bending energy becomes negligible, as shown in \cref{fig:bending_en}.
The peeling angle $\theta$ decreases and the force $F$ increases, as shown in \cref{fig:model_scaling}a,d respectively.
The force required for lifting increases as it needs to compensate for both the tail spring elongation and the loss of adsorbed energy.
This behaviour is radically different from the limiting case of Kendall and Gigli, where both force and angle remain constant throughout the peeling.
Away from the crossover point, the behaviour of the $F$ and $\theta$ becomes increasingly regular, approaching the power-law behaviour marked by dotted lines in \cref{fig:model_scaling}a,d.

In the tethering steady state regime, where the bending contribution $E_\mathrm{bend}$ becomes irrelevant, we can simplify by assuming $B=0$ in \cref{eq:full_energy}.
Moreover, $R$ does not enter the stability equations,  since in tethering regime the curvature is constant, as shown in \cref{fig:model_scaling}b.
The system can thus be described by the geometry in \cref{fig:sketch-bend}c, where two straight segments join at a sharp angle $\theta$.
With this further simplification, we can now drop the inextensibility assumption and  re-introduce the intrinsic elasticity of the GNR.
The total energy
$E(l, \theta; h)$ reads
\begin{align}
\label{eq:en_nobend}
    E = -\epsilon w l + \frac{Q}{2} \left(L - l - \frac{h}{\tan\theta}\right)^2 + \nonumber \\
    \frac{K}{2} \left( L - l - \frac{h}{\sin\theta} \right)^2.
\end{align}
The equilibrium conditions $\partial_\theta E = 0$ and $\partial_l E = 0$ read
\begin{align}
     L-l &= \frac{h}{\sin\theta} \frac{Q\cos\theta + K}{Q+K} - \frac{\epsilon w}{Q+K} \\
     L-l &= \frac{h}{\tan\theta} \frac{Q + K}{Q+K\cos\theta}.
\end{align}
The peeling angle is the solution of a transcendental equation
\begin{align}
    (\tan\theta/2)^2 \left( \tan\theta/2 + \frac{\epsilon w}{2 h} \left( \frac{1}{Q} - \frac{1}{K}\right) \right) = \frac{\epsilon w}{2 h} \left( \frac{1}{Q} + \frac{1}{K}\right) .
\end{align}

In the limit of large height $h \to \infty$, steady-state tethered peeling, the shape of the GNR and the force exerted by the AFM tip approaches a power law
\begin{align}
\label{eq:Tlimth}
    \theta_\mathrm{T} &\approx  2 h^{-1/3} \left(\frac{\epsilon w}{2\widetilde{Q}}\right)^{1/3}  \\
\label{eq:Tliml}
    (L - l)_\mathrm{T} &\approx \frac{h^{4/3}}{2} \left(\frac{2\widetilde{Q}}{\epsilon w}\right)^{1/3} \\
\label{eq:TlimF}
    F_\mathrm{T} &\approx (2 w^2 \epsilon^2 \widetilde{Q} h)^{1/3}
\end{align}
where we introduced the composite stiffness
\begin{equation}
\label{eq:effQ}
    \frac{1}{\widetilde{Q}}=\frac{1}{Q} + \frac{1}{K} .
\end{equation}
The subscript T indicates that the scaling refers to the tethering regime.
This power-law evolution for large $h$ is plotted as dotted lines in \cref{fig:model_scaling}a,b,c,d, colored according to the value of $Q$; in order to compare with the inextensible system described \cref{eq:en_bend}, we take the limit $K \to \infty$ in \cref{eq:effQ}, i.e. $\widetilde{Q}=Q$.
At the considered GNR length $L=\SI{30}{nm}$, the numerical solution approaches the analytic limit for $Q\ge\SI{0.01}{eV/\AA^2}$.
In the limit of infinitely long ribbon length $L\to\infty$, all solutions except that for $Q\to0$ reach the steady regime, see Sec 2 of the SI.
Using the scaling relations \cref{eq:Tliml,eq:Tlimth}, we can estimate when the total energy in \cref{eq:en_nobend} vanishes, obtaining the height $h_\mathrm{detach}$ at which a GNR of length $L$ detaches
\begin{equation}
\label{eq:h_detach}
    h_\mathrm{detach} = \left( \frac{32}{27} \frac{\epsilon w L^3}{Q} \right)^{1/4} .
\end{equation}
This height is indicated by vertical dashed lines the $L-l$ plot, \cref{fig:model_scaling}c.

\subsection{Estimating the regime crossover}
The crossover height at which the regime goes from building the peeling angle to steady state tethering can be estimated analytically.
In the beginning of the lifting $h \to 0$, the bending regime, the evolution of the peeling angle is given by \cref{eq:Blimth,eq:BlimR,eq:BlimF}.
The opposite limit is the tethering regime, $h\to\infty$, as described in \cref{eq:Tlimth,eq:Tliml,eq:TlimF}.
Extrapolating from these limits, we can find the height at which the two solutions for the peeling angle $\theta$ meet $\theta_\mathrm{b}(h_\mathrm{inv})=\theta_\mathrm{t}(h_\mathrm{inv})$.
This point describes the scaling of the crossover height with the systems parameters
\begin{align}
\label{eq:h_inv}
    h_\mathrm{inv} = \left(16\frac{w ^2 B^3}{\epsilon Q^2}\right)^{1/5} \propto Q^{-2/5}.
\end{align}
These heights are marked in \cref{fig:bending_en,fig:model_scaling}a by vertical dashed lines; the color matches the value of $Q$ of the solid curve as reported in the caption.
The crossover estimate is surprisingly good, consider its strong assumptions made in its derivation.
Importantly, this relation could be used to estimate the tethering strength from the crossover point in experimental data.

\section{Discussion}
The suggestive scaling revealed by the analytical model is due to the combined evolution of two points, the bent detachment front, and the tethered tail.
As the tip is vertically lifted, the detachment front and the tail both move towards the center of the adsorbed part, consuming it in a super-linear fashion $l\propto h^{4/3}$.
For the same reason, the force the AFM tip needs to exert to continue the peeling is not constant as in Kendall's and Gigli's limiting cases, but increases sub-linearly $F\propto h^{1/3}$.

These results are obtained in a 2D continuum model which neglects many aspects of the real AFM system.
In order to test the validity of various assumptions made in this approach, we perform realistic MD simulations of GNR on Au(111).
We follow the computational setup in Ref.~\cite{Gigli2019}, which was shown to reproduce experiments, at least qualitatively.
The details of the MD protocol are reported in the Methods section.
These atomistic simulations include many of the ingredients neglected in the model.
The discrete nature of the interface is correctly described, leading to a stick-slip motion rather then a smooth sliding during the detachment dynamics.
As was demonstrated in Ref.~\cite{Gigli2019}, the stick-slip is caused by pinning of both the tail end and the detachment front on the gold substrate corrugation. While very visible experimentally \cite{Kawai2016,Pawlak2020,Gigli2019a}, this stick-slip is weak, and in future peeling of tethered GNRs, will not conceal the simple underlying laws just derived.
The large, but finite, in-plane stiffness allows the GNR able to stretch, relaxing the assumption of the inextensibility constraint also at the beginning of the peeling.
Finally, the AFM tip-GNR anchoring is more realistically modelled by a large, yet finite spring of constant $A > Q,K $.

\begin{figure}
    \centering
    \includegraphics[width=\columnwidth]{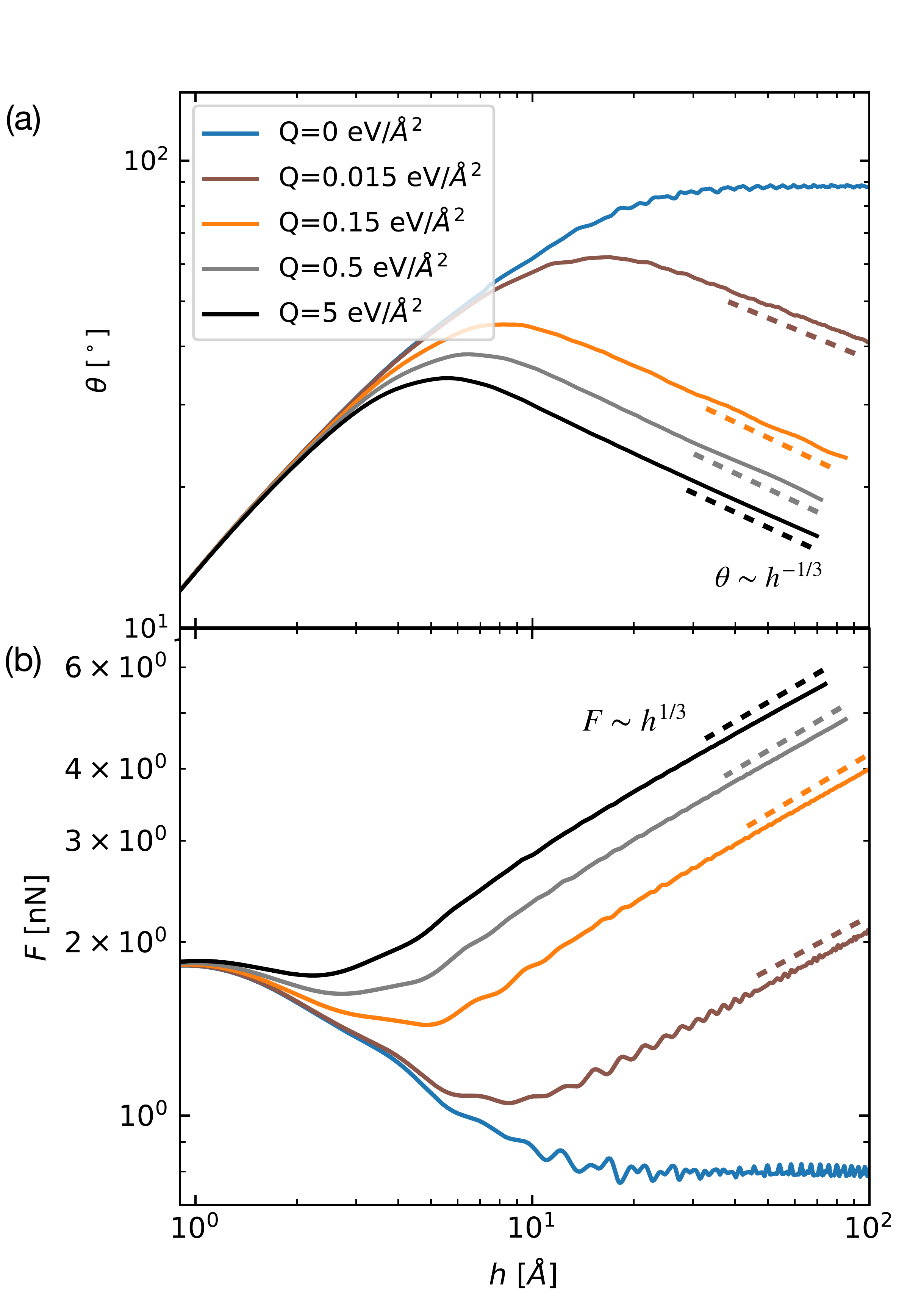}
    \caption{
    (a) Peeling angle $\theta$ computed from MD trajectories as a function of the tip height $h$.
    (b) Force acting on the tip F$_k$ as a function of the tip height $h$.
    The force is computed as described Method.
    Data is plotted in log-log scale.
    }
    \label{fig:MD_thFofh}
\end{figure}
\Cref{fig:MD_thFofh}a reports the evolution of the peeling angle $\theta$ as a function of tip height $h$ for different values of tethering springs $Q$.
The protocol to estimate the peeling angle $\theta$ in MD is presented in Section III of the SI.
The behaviour follows perfectly the prediction of the model: at the beginning the peeling angle increases, independently of the tethering spring, reaches a maximum, whose position decreased with increasing $Q$, and decreases as $\theta\propto h^{-1/3}$, as marked by the dashed lines.
For a dynamic picture of the peeling process see Supplementary Movie 1.
\Cref{fig:MD_thFofh}b reports the evolution of the force $F$ with $h$.
The predicted $F \propto h^{1/3}$ scaling in the steady-state regime is clear in all curves.
Note that the curves are not smooth but, especially at soft tethering, present periodic oscillations.
These oscillations are due to the sliding of the moir\'e units travelling toward the peeling front and detaching\cite{Gigli2019}.
Reasonably, the amplitude of these oscillations decreases with increasing $Q$: the tail tethering strains the GNR and suppresses its rippling, leading to a fainting fingerprint of the moir\'e motion.

These numerical results suggest that our analytical model includes all the ingredients needed to capture the peeling mechanism.
A sizeable unknown in a possible comparison with experiments is the nature and strength of the tethering, which has not been characterized.
Nonetheless, our MD results show that even for finite size GNR ($L=\SI{30}{nm}$) the asymptotic emerges in the measured force, even though the exact predicted exponent of ${1/3}$ is not reached at this finite length.

\begin{figure}
    \centering
    \includegraphics[width=\columnwidth]{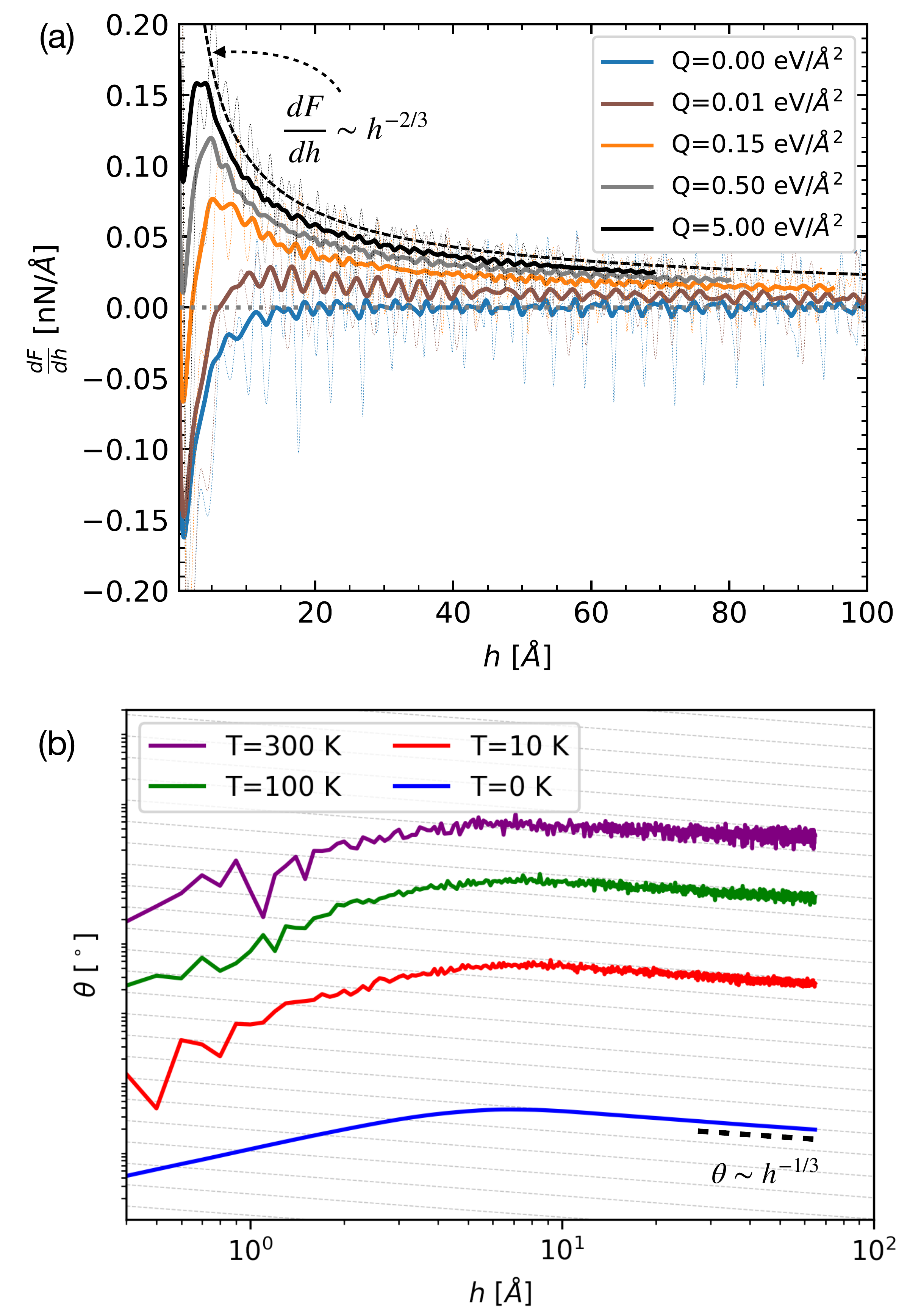}
    \caption{
    Experimental frequency shift as derivative of the simulated AFM force in \cref{fig:MD_thFofh}b.
    Different color refer to different values of $Q$ as in \cref{fig:MD_thFofh}.
    Thin dotted lines report raw value of the numerical derivative, solid lines are running averages of 500 snapshots.
    The black dashed lines sketches the $-2/3$ decay, as indicated by the note.
    See Section IV of the SI for additional details on the finite temperature results.
    }
    \label{fig:MD_dFdh_th}
\end{figure}
A quantity frequently measured in experiments\cite{Gigli2019a} is the shift of the resonant frequency $f$ of the cantilever, which is proportional to the variation of the force.
According to \cref{eq:TlimF}, the frequency shift in the tethered regime scales as $\delta f \propto dF/dh \sim h^{-2/3}$.
The frequency shift is reported in \cref{fig:MD_dFdh_th}a, obtained as a running average of the numerical derivative of the force in \cref{fig:MD_thFofh}.
The signature of the tethering in this observable is the peak emerging at small height in \cref{fig:MD_dFdh_th}a, which becomes more pronounced as the stiffness $Q$ increases.
After this peak, all curves decay with the power-law of -2/3, predicted by the analytic model, and shown in \cref{fig:MD_dFdh_th}a by the dashed black line.

An indirect way to characterise the tethering strength in experiments could be to use  the relationship between the detachment height $h_\mathrm{detach}$, the length $L$ and the tethering strength $Q$.
As the length of the GNR is known and the detachment height can be obtained from the force signal, the tethering strength $Q$ could be estimated from  \cref{eq:h_detach}.

\section{Conclusions}
We report an unexpected and intriguing critical behaviour in the peeling of tethered GNR.
Surprisingly the system behaves radically differently from the two known limits of free tail \cite{Gigli2019a} and immobile tail\cite{Kendall1975}.
Our analytic model predicts a crossover between a transient regime, characterised by the build up of the peeling angle from a flat GNR, to a steady state peeling, governed by non-trivial power law.
Our analytic estimations give a clear picture of these regimes and describe remarkably well the numerical solution.

Our results are in agreement with realistic MD simulations, where the atomistic nature of the contact and the elasticity of the GNR are taken into account.
Hence, we are confident that this scaling should be observable in AFM force traces.

We should mention here that the precise power laws derived depend on the assumption that the ribbon tail tethering can be assimilated to a harmonic spring.
Should the spring be anharmonic and behave otherwise with stretching, that will alter the power laws.
Future experimental peeling results could actually be used to extract the precise stress-strain characteristics of the tethering string.
Whatever the outcome, the general picture, with two regimes separated by a crossover peeling height should be generally valid.

Additional theoretical questions concerning how temperature and quantum effects might alter the classical zero temperature results just derived are postponed to subsequent work, also because current experiments are conducted under cryogenic conditions.
As reported in \cref{fig:MD_dFdh_th}b preliminary finite temperature MD simulations nevertheless suggest that the effects described here will survive qualitatively well beyond cryogenic temperature and even close to room temperature.
On the other hand, quantum effects can be expected, qualitatively inspired by results in graphene \cite{Hasik2018}, to be irrelevant for GNRs at experimental conditions.

Summarizing, a better understanding of the peeling problem on the nanoscale is a useful tool in the hands of the novel field of nano-engineering, where an extreme level of control and precision is required in assemble nanoscale machines and desirable for nanopositioning applications \cite{Ouyang2021,Xue2022,Wei2021}.
By considering the broader class of two-dimensional layered crystals as model systems, these results may also provide useful insights into the tearing and cracking mechanisms of highly confined nanomaterials deposited on substrates, where clear signatures underpin the conversion of bending energy into surface energy of fracture and adhesion\cite{Annett2016PeeliGNR,Hamm2008PeelGNR,Sen2010PeelGNR}.
Besides, from a statistical mechanics point of view, this system represents an interesting example of critical nanomechanical behaviour arising in a simple and well-known system.

\section*{Methods}
All MD simulations were performed using LAMMPS\cite{Plimpton1995,LAMMPS22}.
C-H interactions were modelled using AIREBO potential \cite{Brenner2002}.
Following Ref. \cite{Gigli2018}, C-Au and H-Au interactions were modelled using Lennard-Jones potential with parameters $(\epsilon, \sigma)=(\SI{0.0080}{eV}, \SI{3.42}{\AA})$ and $(\epsilon, \sigma)=(\SI{0.0032}{eV}, \SI{3.42}{\AA})$, respectively.
A cutoff of $\SI{10}{\AA}$ is applied to the LJ potential.
The Au(111) surface is modelled by a Au rigid layer, providing the substrate potential for the GNR peeling.
The GNR of width $w=\SI{7.2}{\AA}$ evolves according to a Langevin dynamics with damping $\gamma=\SI{1}{ps}$ and time-step dt=$\SI{1}{fs}$.
The simulation box is $L_x=\SI{19.979513}{\AA}$ $L_y=\SI{322.984889}{\AA}$ with periodic boundary conditions along $x$ and $y$.

To induce the peeling, the first three leading carbon atoms of the GNR are connected, in the $z$ direction only, to dummy atoms via a spring of constant $A=\SI{1.12356e2}{eV/\AA^2}$ moving at a constant speed $v_\mathrm{lift}=\SI{0.005}{\AA/ps}$ along the $z$ axis.
To create the tethering, the last three trailing carbon atoms of the GNR are connected in all directions to dummy atoms with variable spring constant $Q$ (as reported in the main text), fixed at the initial position of the fully adsorbed GNR.

Force traces are computed as $F=A\frac{1}{3}\sum_{i=0}^{3} (z_i-v_\mathrm{lift}t)$, where $A$ is the tip-GNR bond, where the index $i$ runs over the head atoms and $v_\mathrm{lift}t$ is the position of the dummy atom connected to each \cite{Gigli2019a}.


\begin{acknowledgments}
The authors thank L. Gigli (EPFL), R\'emy Pawlak and Ernst Meyer (University of Basel) for many helpful discussions.
The authors acknowledge support by the Italian Ministry of University and Research through PRIN UTFROM N. 20178PZCB5, the European Union’s Horizon 2020 research and innovation programme under grant agreement No. 899285, and European Union’s H2020 Framework Programme/ERC Advanced Grant N. 8344023 ULTRADISS.
\end{acknowledgments}

\section*{Conflict of Interest}
The authors declare no conflict of interest.

\section*{Authors Contribution}
AV and ET conceptualised the work in the first place and supervised the work.
AS performed the simulations, data analysis and derived the analytical model.
All authors interpreted the results and wrote the manuscript.



\begin{thebibliography}{24}%
\makeatletter
\providecommand \@ifxundefined [1]{%
 \@ifx{#1\undefined}
}%
\providecommand \@ifnum [1]{%
 \ifnum #1\expandafter \@firstoftwo
 \else \expandafter \@secondoftwo
 \fi
}%
\providecommand \@ifx [1]{%
 \ifx #1\expandafter \@firstoftwo
 \else \expandafter \@secondoftwo
 \fi
}%
\providecommand \natexlab [1]{#1}%
\providecommand \enquote  [1]{``#1''}%
\providecommand \bibnamefont  [1]{#1}%
\providecommand \bibfnamefont [1]{#1}%
\providecommand \citenamefont [1]{#1}%
\providecommand \href@noop [0]{\@secondoftwo}%
\providecommand \href [0]{\begingroup \@sanitize@url \@href}%
\providecommand \@href[1]{\@@startlink{#1}\@@href}%
\providecommand \@@href[1]{\endgroup#1\@@endlink}%
\providecommand \@sanitize@url [0]{\catcode `\\12\catcode `\$12\catcode
  `\&12\catcode `\#12\catcode `\^12\catcode `\_12\catcode `\%12\relax}%
\providecommand \@@startlink[1]{}%
\providecommand \@@endlink[0]{}%
\providecommand \url  [0]{\begingroup\@sanitize@url \@url }%
\providecommand \@url [1]{\endgroup\@href {#1}{\urlprefix }}%
\providecommand \urlprefix  [0]{URL }%
\providecommand \Eprint [0]{\href }%
\providecommand \doibase [0]{https://doi.org/}%
\providecommand \selectlanguage [0]{\@gobble}%
\providecommand \bibinfo  [0]{\@secondoftwo}%
\providecommand \bibfield  [0]{\@secondoftwo}%
\providecommand \translation [1]{[#1]}%
\providecommand \BibitemOpen [0]{}%
\providecommand \bibitemStop [0]{}%
\providecommand \bibitemNoStop [0]{.\EOS\space}%
\providecommand \EOS [0]{\spacefactor3000\relax}%
\providecommand \BibitemShut  [1]{\csname bibitem#1\endcsname}%
\let\auto@bib@innerbib\@empty
\bibitem [{\citenamefont {Kendall}(1971)}]{Kendall1971}%
  \BibitemOpen
  \bibfield  {author} {\bibinfo {author} {\bibfnamefont {K.}~\bibnamefont
  {Kendall}},\ }\bibfield  {title} {\bibinfo {title} {{The adhesion and surface
  energy of elastic solids}},\ }\href
  {https://doi.org/10.1088/0022-3727/4/8/320} {\bibfield  {journal} {\bibinfo
  {journal} {Journal of Physics D: Applied Physics}\ }\textbf {\bibinfo
  {volume} {4}},\ \bibinfo {pages} {1186} (\bibinfo {year} {1971})}\BibitemShut
  {NoStop}%
\bibitem [{\citenamefont {Kendall}(1975)}]{Kendall1975}%
  \BibitemOpen
  \bibfield  {author} {\bibinfo {author} {\bibfnamefont {K.}~\bibnamefont
  {Kendall}},\ }\bibfield  {title} {\bibinfo {title} {{Thin-film peeling-the
  elastic term}},\ }\href {https://doi.org/10.1088/0022-3727/8/13/005}
  {\bibfield  {journal} {\bibinfo  {journal} {Journal of Physics D: Applied
  Physics}\ }\textbf {\bibinfo {volume} {8}},\ \bibinfo {pages} {1449}
  (\bibinfo {year} {1975})}\BibitemShut {NoStop}%
\bibitem [{\citenamefont {Manohar}\ \emph {et~al.}(2008)\citenamefont
  {Manohar}, \citenamefont {Mantz}, \citenamefont {Bancroft}, \citenamefont
  {Hui}, \citenamefont {Jagota},\ and\ \citenamefont {Vezenov}}]{Manohar2008}%
  \BibitemOpen
  \bibfield  {author} {\bibinfo {author} {\bibfnamefont {S.}~\bibnamefont
  {Manohar}}, \bibinfo {author} {\bibfnamefont {A.~R.}\ \bibnamefont {Mantz}},
  \bibinfo {author} {\bibfnamefont {K.~E.}\ \bibnamefont {Bancroft}}, \bibinfo
  {author} {\bibfnamefont {C.~Y.}\ \bibnamefont {Hui}}, \bibinfo {author}
  {\bibfnamefont {A.}~\bibnamefont {Jagota}},\ and\ \bibinfo {author}
  {\bibfnamefont {D.~V.}\ \bibnamefont {Vezenov}},\ }\bibfield  {title}
  {\bibinfo {title} {{Peeling single-stranded DNA from graphite surface to
  determine oligonucleotide binding energy by force spectroscopy}},\ }\href
  {https://doi.org/10.1021/nl8022143} {\bibfield  {journal} {\bibinfo
  {journal} {Nano Letters}\ }\textbf {\bibinfo {volume} {8}},\ \bibinfo {pages}
  {4365} (\bibinfo {year} {2008})}\BibitemShut {NoStop}%
\bibitem [{\citenamefont {Vilhena}\ \emph {et~al.}(2018)\citenamefont
  {Vilhena}, \citenamefont {Gnecco}, \citenamefont {Pawlak}, \citenamefont
  {Moreno-Herrero}, \citenamefont {Meyer},\ and\ \citenamefont
  {P{\'{e}}rez}}]{Vilhena2018}%
  \BibitemOpen
  \bibfield  {author} {\bibinfo {author} {\bibfnamefont {J.~G.}\ \bibnamefont
  {Vilhena}}, \bibinfo {author} {\bibfnamefont {E.}~\bibnamefont {Gnecco}},
  \bibinfo {author} {\bibfnamefont {R.}~\bibnamefont {Pawlak}}, \bibinfo
  {author} {\bibfnamefont {F.}~\bibnamefont {Moreno-Herrero}}, \bibinfo
  {author} {\bibfnamefont {E.}~\bibnamefont {Meyer}},\ and\ \bibinfo {author}
  {\bibfnamefont {R.}~\bibnamefont {P{\'{e}}rez}},\ }\bibfield  {title}
  {\bibinfo {title} {{Stick-Slip Motion of ssDNA over Graphene}},\ }\href
  {https://doi.org/10.1021/acs.jpcb.7b06952} {\bibfield  {journal} {\bibinfo
  {journal} {Journal of Physical Chemistry B}\ }\textbf {\bibinfo {volume}
  {122}},\ \bibinfo {pages} {840} (\bibinfo {year} {2018})}\BibitemShut
  {NoStop}%
\bibitem [{\citenamefont {Kawai}\ \emph {et~al.}(2016)\citenamefont {Kawai},
  \citenamefont {Benassi}, \citenamefont {Gnecco}, \citenamefont {S{\"{o}}de},
  \citenamefont {Pawlak}, \citenamefont {Feng}, \citenamefont {M{\"{u}}llen},
  \citenamefont {Passerone}, \citenamefont {Pignedoli}, \citenamefont
  {Ruffieux}, \citenamefont {Fasel},\ and\ \citenamefont {Meyer}}]{Kawai2016}%
  \BibitemOpen
  \bibfield  {author} {\bibinfo {author} {\bibfnamefont {S.}~\bibnamefont
  {Kawai}}, \bibinfo {author} {\bibfnamefont {A.}~\bibnamefont {Benassi}},
  \bibinfo {author} {\bibfnamefont {E.}~\bibnamefont {Gnecco}}, \bibinfo
  {author} {\bibfnamefont {H.}~\bibnamefont {S{\"{o}}de}}, \bibinfo {author}
  {\bibfnamefont {R.}~\bibnamefont {Pawlak}}, \bibinfo {author} {\bibfnamefont
  {X.}~\bibnamefont {Feng}}, \bibinfo {author} {\bibfnamefont {K.}~\bibnamefont
  {M{\"{u}}llen}}, \bibinfo {author} {\bibfnamefont {D.}~\bibnamefont
  {Passerone}}, \bibinfo {author} {\bibfnamefont {C.~A.}\ \bibnamefont
  {Pignedoli}}, \bibinfo {author} {\bibfnamefont {P.}~\bibnamefont {Ruffieux}},
  \bibinfo {author} {\bibfnamefont {R.}~\bibnamefont {Fasel}},\ and\ \bibinfo
  {author} {\bibfnamefont {E.}~\bibnamefont {Meyer}},\ }\bibfield  {title}
  {\bibinfo {title} {{Superlubricity of graphene nanoribbons on gold
  surfaces}},\ }\href {https://doi.org/10.1126/science.aad3569} {\bibfield
  {journal} {\bibinfo  {journal} {Science}\ }\textbf {\bibinfo {volume}
  {351}},\ \bibinfo {pages} {957} (\bibinfo {year} {2016})}\BibitemShut
  {NoStop}%
\bibitem [{\citenamefont {Pawlak}\ \emph {et~al.}(2020)\citenamefont {Pawlak},
  \citenamefont {Vilhena}, \citenamefont {D'astolfo}, \citenamefont {Liu},
  \citenamefont {Prampolini}, \citenamefont {Meier}, \citenamefont {Glatzel},
  \citenamefont {Lemkul}, \citenamefont {H{\"{a}}ner}, \citenamefont
  {Decurtins}, \citenamefont {Baratoff}, \citenamefont {P{\'{e}}rez},
  \citenamefont {Liu},\ and\ \citenamefont {Meyer}}]{Pawlak2020}%
  \BibitemOpen
  \bibfield  {author} {\bibinfo {author} {\bibfnamefont {R.}~\bibnamefont
  {Pawlak}}, \bibinfo {author} {\bibfnamefont {J.~G.}\ \bibnamefont {Vilhena}},
  \bibinfo {author} {\bibfnamefont {P.}~\bibnamefont {D'astolfo}}, \bibinfo
  {author} {\bibfnamefont {X.}~\bibnamefont {Liu}}, \bibinfo {author}
  {\bibfnamefont {G.}~\bibnamefont {Prampolini}}, \bibinfo {author}
  {\bibfnamefont {T.}~\bibnamefont {Meier}}, \bibinfo {author} {\bibfnamefont
  {T.}~\bibnamefont {Glatzel}}, \bibinfo {author} {\bibfnamefont {J.~A.}\
  \bibnamefont {Lemkul}}, \bibinfo {author} {\bibfnamefont {R.}~\bibnamefont
  {H{\"{a}}ner}}, \bibinfo {author} {\bibfnamefont {S.}~\bibnamefont
  {Decurtins}}, \bibinfo {author} {\bibfnamefont {A.}~\bibnamefont {Baratoff}},
  \bibinfo {author} {\bibfnamefont {R.}~\bibnamefont {P{\'{e}}rez}}, \bibinfo
  {author} {\bibfnamefont {S.~X.}\ \bibnamefont {Liu}},\ and\ \bibinfo {author}
  {\bibfnamefont {E.}~\bibnamefont {Meyer}},\ }\bibfield  {title} {\bibinfo
  {title} {{Sequential Bending and Twisting around C-C Single Bonds by
  Mechanical Lifting of a Pre-Adsorbed Polymer}},\ }\href
  {https://doi.org/10.1021/acs.nanolett.9b04418} {\bibfield  {journal}
  {\bibinfo  {journal} {Nano Letters}\ }\textbf {\bibinfo {volume} {20}},\
  \bibinfo {pages} {652} (\bibinfo {year} {2020})}\BibitemShut {NoStop}%
\bibitem [{\citenamefont {Gigli}\ \emph
  {et~al.}(2019{\natexlab{a}})\citenamefont {Gigli}, \citenamefont {Kawai},
  \citenamefont {Guerra}, \citenamefont {Manini}, \citenamefont {Pawlak},
  \citenamefont {Feng}, \citenamefont {M{\"{u}}llen}, \citenamefont {Ruffieux},
  \citenamefont {Fasel}, \citenamefont {Tosatti}, \citenamefont {Meyer},\ and\
  \citenamefont {Vanossi}}]{Gigli2019a}%
  \BibitemOpen
  \bibfield  {author} {\bibinfo {author} {\bibfnamefont {L.}~\bibnamefont
  {Gigli}}, \bibinfo {author} {\bibfnamefont {S.}~\bibnamefont {Kawai}},
  \bibinfo {author} {\bibfnamefont {R.}~\bibnamefont {Guerra}}, \bibinfo
  {author} {\bibfnamefont {N.}~\bibnamefont {Manini}}, \bibinfo {author}
  {\bibfnamefont {R.}~\bibnamefont {Pawlak}}, \bibinfo {author} {\bibfnamefont
  {X.}~\bibnamefont {Feng}}, \bibinfo {author} {\bibfnamefont {K.}~\bibnamefont
  {M{\"{u}}llen}}, \bibinfo {author} {\bibfnamefont {P.}~\bibnamefont
  {Ruffieux}}, \bibinfo {author} {\bibfnamefont {R.}~\bibnamefont {Fasel}},
  \bibinfo {author} {\bibfnamefont {E.}~\bibnamefont {Tosatti}}, \bibinfo
  {author} {\bibfnamefont {E.}~\bibnamefont {Meyer}},\ and\ \bibinfo {author}
  {\bibfnamefont {A.}~\bibnamefont {Vanossi}},\ }\bibfield  {title} {\bibinfo
  {title} {{Detachment Dynamics of Graphene Nanoribbons on Gold}},\ }\href
  {https://doi.org/10.1021/acsnano.8b07894} {\bibfield  {journal} {\bibinfo
  {journal} {ACS Nano}\ }\textbf {\bibinfo {volume} {13}},\ \bibinfo {pages}
  {689} (\bibinfo {year} {2019}{\natexlab{a}})}\BibitemShut {NoStop}%
\bibitem [{\citenamefont {Vanossi}\ \emph {et~al.}(2020)\citenamefont
  {Vanossi}, \citenamefont {Bechinger},\ and\ \citenamefont
  {Urbakh}}]{Vanossi2020a}%
  \BibitemOpen
  \bibfield  {author} {\bibinfo {author} {\bibfnamefont {A.}~\bibnamefont
  {Vanossi}}, \bibinfo {author} {\bibfnamefont {C.}~\bibnamefont {Bechinger}},\
  and\ \bibinfo {author} {\bibfnamefont {M.}~\bibnamefont {Urbakh}},\
  }\bibfield  {title} {\bibinfo {title} {{Structural lubricity in soft and hard
  matter systems}},\ }\href {https://doi.org/10.1038/s41467-020-18429-1}
  {\bibfield  {journal} {\bibinfo  {journal} {Nature Communications}\ }\textbf
  {\bibinfo {volume} {11}},\ \bibinfo {pages} {4657} (\bibinfo {year}
  {2020})}\BibitemShut {NoStop}%
\bibitem [{\citenamefont {Gigli}\ \emph {et~al.}(2017)\citenamefont {Gigli},
  \citenamefont {Manini}, \citenamefont {Benassi}, \citenamefont {Tosatti},
  \citenamefont {Vanossi},\ and\ \citenamefont {Guerra}}]{Gigli2017}%
  \BibitemOpen
  \bibfield  {author} {\bibinfo {author} {\bibfnamefont {L.}~\bibnamefont
  {Gigli}}, \bibinfo {author} {\bibfnamefont {N.}~\bibnamefont {Manini}},
  \bibinfo {author} {\bibfnamefont {A.}~\bibnamefont {Benassi}}, \bibinfo
  {author} {\bibfnamefont {E.}~\bibnamefont {Tosatti}}, \bibinfo {author}
  {\bibfnamefont {A.}~\bibnamefont {Vanossi}},\ and\ \bibinfo {author}
  {\bibfnamefont {R.}~\bibnamefont {Guerra}},\ }\bibfield  {title} {\bibinfo
  {title} {{Graphene nanoribbons on gold: Understanding superlubricity and edge
  effects}},\ }\href {https://doi.org/10.1088/2053-1583/aa7fdf} {\bibfield
  {journal} {\bibinfo  {journal} {2D Materials}\ }\textbf {\bibinfo {volume}
  {4}},\ \bibinfo {pages} {045003} (\bibinfo {year} {2017})}\BibitemShut
  {NoStop}%
\bibitem [{\citenamefont {Vanossi}\ \emph {et~al.}(2013)\citenamefont
  {Vanossi}, \citenamefont {Manini}, \citenamefont {Urbakh}, \citenamefont
  {Zapperi},\ and\ \citenamefont {Tosatti}}]{Vanossi2013}%
  \BibitemOpen
  \bibfield  {author} {\bibinfo {author} {\bibfnamefont {A.}~\bibnamefont
  {Vanossi}}, \bibinfo {author} {\bibfnamefont {N.}~\bibnamefont {Manini}},
  \bibinfo {author} {\bibfnamefont {M.}~\bibnamefont {Urbakh}}, \bibinfo
  {author} {\bibfnamefont {S.}~\bibnamefont {Zapperi}},\ and\ \bibinfo {author}
  {\bibfnamefont {E.}~\bibnamefont {Tosatti}},\ }\bibfield  {title} {\bibinfo
  {title} {{Colloquium: Modeling friction: From nanoscale to mesoscale}},\
  }\href {https://doi.org/10.1103/RevModPhys.85.529} {\bibfield  {journal}
  {\bibinfo  {journal} {Reviews of Modern Physics}\ }\textbf {\bibinfo {volume}
  {85}},\ \bibinfo {pages} {529} (\bibinfo {year} {2013})}\BibitemShut
  {NoStop}%
\bibitem [{\citenamefont {Gigli}\ \emph {et~al.}(2018)\citenamefont {Gigli},
  \citenamefont {Manini}, \citenamefont {Tosatti}, \citenamefont {Guerra},\
  and\ \citenamefont {Vanossi}}]{Gigli2018}%
  \BibitemOpen
  \bibfield  {author} {\bibinfo {author} {\bibfnamefont {L.}~\bibnamefont
  {Gigli}}, \bibinfo {author} {\bibfnamefont {N.}~\bibnamefont {Manini}},
  \bibinfo {author} {\bibfnamefont {E.}~\bibnamefont {Tosatti}}, \bibinfo
  {author} {\bibfnamefont {R.}~\bibnamefont {Guerra}},\ and\ \bibinfo {author}
  {\bibfnamefont {A.}~\bibnamefont {Vanossi}},\ }\bibfield  {title} {\bibinfo
  {title} {{Lifted graphene nanoribbons on gold: From smooth sliding to
  multiple stick-slip regimes}},\ }\href {https://doi.org/10.1039/c7nr07857a}
  {\bibfield  {journal} {\bibinfo  {journal} {Nanoscale}\ }\textbf {\bibinfo
  {volume} {10}},\ \bibinfo {pages} {2073} (\bibinfo {year}
  {2018})}\BibitemShut {NoStop}%
\bibitem [{\citenamefont {Gigli}\ \emph
  {et~al.}(2019{\natexlab{b}})\citenamefont {Gigli}, \citenamefont {Vanossi},\
  and\ \citenamefont {Tosatti}}]{Gigli2019}%
  \BibitemOpen
  \bibfield  {author} {\bibinfo {author} {\bibfnamefont {L.}~\bibnamefont
  {Gigli}}, \bibinfo {author} {\bibfnamefont {A.}~\bibnamefont {Vanossi}},\
  and\ \bibinfo {author} {\bibfnamefont {E.}~\bibnamefont {Tosatti}},\
  }\bibfield  {title} {\bibinfo {title} {{Modeling nanoribbon peeling}},\
  }\href {https://doi.org/10.1039/c9nr04821a} {\bibfield  {journal} {\bibinfo
  {journal} {Nanoscale}\ }\textbf {\bibinfo {volume} {11}},\ \bibinfo {pages}
  {17396} (\bibinfo {year} {2019}{\natexlab{b}})}\BibitemShut {NoStop}%
\bibitem [{\citenamefont {Bosak}\ \emph {et~al.}(2007)\citenamefont {Bosak},
  \citenamefont {Krisch}, \citenamefont {Mohr}, \citenamefont {Maultzsch},\
  and\ \citenamefont {Thomsen}}]{Bosak2007G_elasticExp}%
  \BibitemOpen
  \bibfield  {author} {\bibinfo {author} {\bibfnamefont {A.}~\bibnamefont
  {Bosak}}, \bibinfo {author} {\bibfnamefont {M.}~\bibnamefont {Krisch}},
  \bibinfo {author} {\bibfnamefont {M.}~\bibnamefont {Mohr}}, \bibinfo {author}
  {\bibfnamefont {J.}~\bibnamefont {Maultzsch}},\ and\ \bibinfo {author}
  {\bibfnamefont {C.}~\bibnamefont {Thomsen}},\ }\bibfield  {title} {\bibinfo
  {title} {{Elasticity of single-crystalline graphite: Inelastic x-ray
  scattering study}},\ }\href {https://doi.org/10.1103/PhysRevB.75.153408}
  {\bibfield  {journal} {\bibinfo  {journal} {Physical Review B}\ }\textbf
  {\bibinfo {volume} {75}},\ \bibinfo {pages} {153408} (\bibinfo {year}
  {2007})}\BibitemShut {NoStop}%
\bibitem [{\citenamefont {Andrew}\ \emph {et~al.}(2012)\citenamefont {Andrew},
  \citenamefont {Mapasha}, \citenamefont {Ukpong},\ and\ \citenamefont
  {Chetty}}]{Andrew2012G_elastic}%
  \BibitemOpen
  \bibfield  {author} {\bibinfo {author} {\bibfnamefont {R.~C.}\ \bibnamefont
  {Andrew}}, \bibinfo {author} {\bibfnamefont {R.~E.}\ \bibnamefont {Mapasha}},
  \bibinfo {author} {\bibfnamefont {A.~M.}\ \bibnamefont {Ukpong}},\ and\
  \bibinfo {author} {\bibfnamefont {N.}~\bibnamefont {Chetty}},\ }\bibfield
  {title} {\bibinfo {title} {{Mechanical properties of graphene and
  boronitrene}},\ }\href {https://doi.org/10.1103/PhysRevB.85.125428}
  {\bibfield  {journal} {\bibinfo  {journal} {Physical Review B}\ }\textbf
  {\bibinfo {volume} {85}},\ \bibinfo {pages} {125428} (\bibinfo {year}
  {2012})}\BibitemShut {NoStop}%
\bibitem [{\citenamefont {Ha{\v{s}}{\'{i}}k}\ \emph {et~al.}(2018)\citenamefont
  {Ha{\v{s}}{\'{i}}k}, \citenamefont {Tosatti},\ and\ \citenamefont
  {Martoň{\'{a}}k}}]{Hasik2018}%
  \BibitemOpen
  \bibfield  {author} {\bibinfo {author} {\bibfnamefont {J.}~\bibnamefont
  {Ha{\v{s}}{\'{i}}k}}, \bibinfo {author} {\bibfnamefont {E.}~\bibnamefont
  {Tosatti}},\ and\ \bibinfo {author} {\bibfnamefont {R.}~\bibnamefont
  {Martoň{\'{a}}k}},\ }\bibfield  {title} {\bibinfo {title} {{Quantum and
  classical ripples in graphene}},\ }\href
  {https://doi.org/10.1103/PhysRevB.97.140301} {\bibfield  {journal} {\bibinfo
  {journal} {Physical Review B}\ }\textbf {\bibinfo {volume} {97}},\ \bibinfo
  {pages} {3} (\bibinfo {year} {2018})}\BibitemShut {NoStop}%
\bibitem [{\citenamefont {Ouyang}\ \emph {et~al.}(2021)\citenamefont {Ouyang},
  \citenamefont {Hod},\ and\ \citenamefont {Urbakh}}]{Ouyang2021}%
  \BibitemOpen
  \bibfield  {author} {\bibinfo {author} {\bibfnamefont {W.}~\bibnamefont
  {Ouyang}}, \bibinfo {author} {\bibfnamefont {O.}~\bibnamefont {Hod}},\ and\
  \bibinfo {author} {\bibfnamefont {M.}~\bibnamefont {Urbakh}},\ }\bibfield
  {title} {\bibinfo {title} {{Registry-Dependent Peeling of Layered Material
  Interfaces: The Case of Graphene Nanoribbons on Hexagonal Boron Nitride}},\
  }\bibfield  {journal} {\bibinfo  {journal} {ACS Applied Materials {\&}
  Interfaces}\ }\href {https://doi.org/10.1021/acsami.1c09529}
  {10.1021/acsami.1c09529} (\bibinfo {year} {2021})\BibitemShut {NoStop}%
\bibitem [{\citenamefont {Xue}\ \emph {et~al.}(2022)\citenamefont {Xue},
  \citenamefont {Chen}, \citenamefont {Wang},\ and\ \citenamefont
  {Huang}}]{Xue2022}%
  \BibitemOpen
  \bibfield  {author} {\bibinfo {author} {\bibfnamefont {Z.}~\bibnamefont
  {Xue}}, \bibinfo {author} {\bibfnamefont {G.}~\bibnamefont {Chen}}, \bibinfo
  {author} {\bibfnamefont {C.}~\bibnamefont {Wang}},\ and\ \bibinfo {author}
  {\bibfnamefont {R.}~\bibnamefont {Huang}},\ }\bibfield  {title} {\bibinfo
  {title} {{Peeling and sliding of graphene nanoribbons with periodic van der
  Waals interactions}},\ }\bibfield  {journal} {\bibinfo  {journal} {Journal of
  the Mechanics and Physics of Solids}\ }\textbf {\bibinfo {volume} {158}},\
  \href {https://doi.org/10.1016/j.jmps.2021.104698}
  {10.1016/j.jmps.2021.104698} (\bibinfo {year} {2022})\BibitemShut {NoStop}%
\bibitem [{\citenamefont {Wei}\ \emph {et~al.}(2021)\citenamefont {Wei},
  \citenamefont {Lin}, \citenamefont {Wang},\ and\ \citenamefont
  {Zhao}}]{Wei2021}%
  \BibitemOpen
  \bibfield  {author} {\bibinfo {author} {\bibfnamefont {Z.~X.}\ \bibnamefont
  {Wei}}, \bibinfo {author} {\bibfnamefont {K.}~\bibnamefont {Lin}}, \bibinfo
  {author} {\bibfnamefont {X.~H.}\ \bibnamefont {Wang}},\ and\ \bibinfo
  {author} {\bibfnamefont {Y.~P.}\ \bibnamefont {Zhao}},\ }\bibfield  {title}
  {\bibinfo {title} {{Peeling of graphene/molybdenum disulfide heterostructure
  at different angles: A continuum model with accommodations for van der Waals
  interaction}},\ }\bibfield  {journal} {\bibinfo  {journal} {Composites Part
  A: Applied Science and Manufacturing}\ }\textbf {\bibinfo {volume} {150}},\
  \href {https://doi.org/10.1016/j.compositesa.2021.106592}
  {10.1016/j.compositesa.2021.106592} (\bibinfo {year} {2021})\BibitemShut
  {NoStop}%
\bibitem [{\citenamefont {Annett}\ and\ \citenamefont
  {Cross}(2016)}]{Annett2016PeeliGNR}%
  \BibitemOpen
  \bibfield  {author} {\bibinfo {author} {\bibfnamefont {J.}~\bibnamefont
  {Annett}}\ and\ \bibinfo {author} {\bibfnamefont {G.~L.~W.}\ \bibnamefont
  {Cross}},\ }\bibfield  {title} {\bibinfo {title} {{Self-assembly of graphene
  ribbons by spontaneous self-tearing and peeling from a substrate}},\ }\href
  {https://doi.org/10.1038/nature18304} {\bibfield  {journal} {\bibinfo
  {journal} {Nature}\ }\textbf {\bibinfo {volume} {535}},\ \bibinfo {pages}
  {271} (\bibinfo {year} {2016})}\BibitemShut {NoStop}%
\bibitem [{\citenamefont {Hamm}\ \emph {et~al.}(2008)\citenamefont {Hamm},
  \citenamefont {Reis}, \citenamefont {LeBlanc}, \citenamefont {Roman},\ and\
  \citenamefont {Cerda}}]{Hamm2008PeelGNR}%
  \BibitemOpen
  \bibfield  {author} {\bibinfo {author} {\bibfnamefont {E.}~\bibnamefont
  {Hamm}}, \bibinfo {author} {\bibfnamefont {P.}~\bibnamefont {Reis}}, \bibinfo
  {author} {\bibfnamefont {M.}~\bibnamefont {LeBlanc}}, \bibinfo {author}
  {\bibfnamefont {B.}~\bibnamefont {Roman}},\ and\ \bibinfo {author}
  {\bibfnamefont {E.}~\bibnamefont {Cerda}},\ }\bibfield  {title} {\bibinfo
  {title} {{Tearing as a test for mechanical characterization of thin adhesive
  films}},\ }\href {https://doi.org/10.1038/nmat2161} {\bibfield  {journal}
  {\bibinfo  {journal} {Nature Materials}\ }\textbf {\bibinfo {volume} {7}},\
  \bibinfo {pages} {386} (\bibinfo {year} {2008})}\BibitemShut {NoStop}%
\bibitem [{\citenamefont {Sen}\ \emph {et~al.}(2010)\citenamefont {Sen},
  \citenamefont {Novoselov}, \citenamefont {Reis},\ and\ \citenamefont
  {Buehler}}]{Sen2010PeelGNR}%
  \BibitemOpen
  \bibfield  {author} {\bibinfo {author} {\bibfnamefont {D.}~\bibnamefont
  {Sen}}, \bibinfo {author} {\bibfnamefont {K.~S.}\ \bibnamefont {Novoselov}},
  \bibinfo {author} {\bibfnamefont {P.~M.}\ \bibnamefont {Reis}},\ and\
  \bibinfo {author} {\bibfnamefont {M.~J.}\ \bibnamefont {Buehler}},\
  }\bibfield  {title} {\bibinfo {title} {{Tearing Graphene Sheets From Adhesive
  Substrates Produces Tapered Nanoribbons}},\ }\href
  {https://doi.org/10.1002/smll.201000097} {\bibfield  {journal} {\bibinfo
  {journal} {Small}\ }\textbf {\bibinfo {volume} {6}},\ \bibinfo {pages} {1108}
  (\bibinfo {year} {2010})}\BibitemShut {NoStop}%
\bibitem [{\citenamefont {Plimpton}(1995)}]{Plimpton1995}%
  \BibitemOpen
  \bibfield  {author} {\bibinfo {author} {\bibfnamefont {S.}~\bibnamefont
  {Plimpton}},\ }\bibfield  {title} {\bibinfo {title} {{Fast parallel
  algorithms for short-range molecular dynamics}},\ }\href
  {https://doi.org/10.1006/jcph.1995.1039} {\bibfield  {journal} {\bibinfo
  {journal} {Journal of Computational Physics}\ }\textbf {\bibinfo {volume}
  {117}},\ \bibinfo {pages} {1} (\bibinfo {year} {1995})}\BibitemShut {NoStop}%
\bibitem [{\citenamefont {Thompson}\ \emph {et~al.}(2022)\citenamefont
  {Thompson}, \citenamefont {Aktulga}, \citenamefont {Berger}, \citenamefont
  {Bolintineanu}, \citenamefont {Brown}, \citenamefont {Crozier}, \citenamefont
  {in~'t Veld}, \citenamefont {Kohlmeyer}, \citenamefont {Moore}, \citenamefont
  {Nguyen}, \citenamefont {Shan}, \citenamefont {Stevens}, \citenamefont
  {Tranchida}, \citenamefont {Trott},\ and\ \citenamefont
  {Plimpton}}]{LAMMPS22}%
  \BibitemOpen
  \bibfield  {author} {\bibinfo {author} {\bibfnamefont {A.~P.}\ \bibnamefont
  {Thompson}}, \bibinfo {author} {\bibfnamefont {H.~M.}\ \bibnamefont
  {Aktulga}}, \bibinfo {author} {\bibfnamefont {R.}~\bibnamefont {Berger}},
  \bibinfo {author} {\bibfnamefont {D.~S.}\ \bibnamefont {Bolintineanu}},
  \bibinfo {author} {\bibfnamefont {W.~M.}\ \bibnamefont {Brown}}, \bibinfo
  {author} {\bibfnamefont {P.~S.}\ \bibnamefont {Crozier}}, \bibinfo {author}
  {\bibfnamefont {P.~J.}\ \bibnamefont {in~'t Veld}}, \bibinfo {author}
  {\bibfnamefont {A.}~\bibnamefont {Kohlmeyer}}, \bibinfo {author}
  {\bibfnamefont {S.~G.}\ \bibnamefont {Moore}}, \bibinfo {author}
  {\bibfnamefont {T.~D.}\ \bibnamefont {Nguyen}}, \bibinfo {author}
  {\bibfnamefont {R.}~\bibnamefont {Shan}}, \bibinfo {author} {\bibfnamefont
  {M.~J.}\ \bibnamefont {Stevens}}, \bibinfo {author} {\bibfnamefont
  {J.}~\bibnamefont {Tranchida}}, \bibinfo {author} {\bibfnamefont
  {C.}~\bibnamefont {Trott}},\ and\ \bibinfo {author} {\bibfnamefont {S.~J.}\
  \bibnamefont {Plimpton}},\ }\bibfield  {title} {\bibinfo {title} {{LAMMPS - a
  flexible simulation tool for particle-based materials modeling at the atomic,
  meso, and continuum scales}},\ }\bibfield  {journal} {\bibinfo  {journal}
  {Computer Physics Communications}\ }\textbf {\bibinfo {volume} {271}},\ \href
  {https://doi.org/10.1016/j.cpc.2021.108171} {10.1016/j.cpc.2021.108171}
  (\bibinfo {year} {2022})\BibitemShut {NoStop}%
\bibitem [{\citenamefont {Brenner}\ \emph {et~al.}(2002)\citenamefont
  {Brenner}, \citenamefont {Shenderova}, \citenamefont {Harrison},
  \citenamefont {Stuart}, \citenamefont {Ni},\ and\ \citenamefont
  {Sinnott}}]{Brenner2002}%
  \BibitemOpen
  \bibfield  {author} {\bibinfo {author} {\bibfnamefont {D.~W.}\ \bibnamefont
  {Brenner}}, \bibinfo {author} {\bibfnamefont {O.~A.}\ \bibnamefont
  {Shenderova}}, \bibinfo {author} {\bibfnamefont {J.~A.}\ \bibnamefont
  {Harrison}}, \bibinfo {author} {\bibfnamefont {S.~J.}\ \bibnamefont
  {Stuart}}, \bibinfo {author} {\bibfnamefont {B.}~\bibnamefont {Ni}},\ and\
  \bibinfo {author} {\bibfnamefont {S.~B.}\ \bibnamefont {Sinnott}},\
  }\bibfield  {title} {\bibinfo {title} {{A second-generation reactive
  empirical bond order (REBO) potential energy expression for hydrocarbons}},\
  }\href {http://stacks.iop.org/0953-8984/14/i=4/a=312} {\bibfield  {journal}
  {\bibinfo  {journal} {Journal of Physics-Condensed Matter}\ }\textbf
  {\bibinfo {volume} {14}},\ \bibinfo {pages} {783} (\bibinfo {year}
  {2002})}\BibitemShut {NoStop}%
\end{thebibliography}
%

\end{document}